\documentclass[12pt,preprint]{aastex}
\usepackage{graphicx}

\shorttitle{N$_{2}$H$^{+}$ Observations of Class 0 protostars}
\shortauthors{Chen et al.}

\begin{document}

\title{OVRO N$_{2}$H$^{+}$ Observations of Class 0 Protostars: Constraints on the Formation of Binary
Stars}

\author{Xuepeng Chen, Ralf Launhardt, and Thomas Henning}
\affil{Max Planck Institute for Astronomy, K\"{o}nigstuhl 17,
D-69117 Heidelberg, Germany; chen@mpia.de}


\begin{abstract}

We present the results of an interferometric study of the
N$_{2}$H$^{+}$\,(1--0) emission from nine nearby, isolated,
low-mass protostellar cores, using the OVRO millimeter array.
The main goal of this study is the kinematic characterization
of the cores in terms of rotation, turbulence, and fragmentation.
Eight of the nine objects have compact N$_{2}$H$^{+}$ cores with 
FWHM radii of 1200 $-$ 3500 AU, spatially coinciding with the 
thermal dust continuum emission. The only more evolved (Class I) 
object in the sample (CB\,188) shows only faint and extended N$_{2}$H$^{+}$\
emission. The mean N$_{2}$H$^{+}$\ line width was found to be 
0.37\,km s$^{-1}$. Estimated virial masses range from 0.3 to 1.2\,$M_\odot$. 
We find that thermal and turbulent energy support are about equally
important in these cores, while rotational support is negligible.
The measured velocity gradients across the cores range from 6 to
24\,km\,s$^{-1}$\,pc$^{-1}$. Assuming these gradients are produced
by bulk rotation, we find that the specific angular momenta of the
observed Class\,0 protostellar cores are intermediate between
those of dense (prestellar) molecular cloud cores and the orbital
angular momenta of wide PMS binary systems. There appears to be no
evolution (decrease) of angular momentum from the smallest
prestellar cores via protostellar cores to wide PMS binary systems.
In the context that most protostellar cores are assumed to
fragment and form binary stars, this means that most of the
angular momentum contained in the collapse region is transformed
into orbital angular momentum of the resulting stellar binary
systems.

\end{abstract}


\section{INTRODUCTION}

A major gap in our understanding of star formation concerns the
origin of binary stars. Binary/multiple systems appear to be the
preferred outcome of the star formation process, but at present we
do not understand how this occurs (Mathieu et al. 2000). In the
past two decades, statistical properties of binary stars are
gradually comprehended through numerous observations and
theoretical simulations, but are still subject to extensive
ongoing studies (see a review by Duch\^{e}ne et al. 2007). The
most recent observational results and theoretical models for
binary star formation can be found in the proceedings of IAU
Symposium 200 (Zinnecker \& Mathieu 2001), Launhardt (2004), and
reviews in Protostars \& Protoplanets V (Reipurth, Jewitt, \& Keil
2007).

Different formation scenarios for binary stars have been proposed,
of which the classical ideas of capture and fission are no longer
considered major processes. Fragmentation of rotating cloud cores
with initially flat density profiles, immediately after a phase of
free-fall collapse, is generally considered to be the most
efficient mechanism, leading to systems with a wide variety of
properties (see reviews by Bodenheimer et al. 2000 and Tohline
2002). These properties are largely determined by the accretion
process which, in turn, strongly depends on the initial conditions
of the cloud cores, e.g., the initial distributions of mass and
angular momentum (see Bate \& Bonnell 1997). Some of the
theoretical predictions, e.g., that close binary systems are
likely to have mass ratios near unity (see Bate 2000), are already
indirectly supported by statistical studies of evolved binary
systems (see Halbwachs et al. 2003).

However, our current knowledge on the formation of binary stars
mainly relies on observations of main sequence (MS) and pre-main
sequence (PMS) stars and the constraints they put on the models.
The observational link between initial conditions in a molecular
cloud and the final star systems formed therein is still missing.
Furthermore, as multiple systems certainly undergo dynamical
evolution, important information about the formation phase is lost
in the final systems. Direct observations of prestellar cores and
protostars are needed therefore to answer a number of questions,
e.g., how common is binarity/multiplicity in the protostellar
phase? What makes a prestellar core to fragment in the collapse
phase? How is angular momentum distributed? Are there differences
between cores forming binaries and those forming single stars?
Unfortunately, direct observations of protostellar stages, when
the main collapse has started but no optical or infrared emission
emerges from the protostar through the opaque infalling envelope,
were long hampered by the low angular resolution of millimeter
(mm) telescopes and the results were mostly interpreted in terms
of single star formation. Only the recent advance of large mm
interferometers has enabled us to directly observe the formation
phase of binary stars, although the number of known systems is
still very small and no systematic observational studies of the
initial fragmentation process do exist yet (see the review by
Launhardt 2004). The systematic study by Looney et al. (2000) was
successful in detecting a number of protostellar binaries by mm
dust continuum emission, but did not provide any kinematic
information. Only recently these authors published velocity fields
of 3 objects from their sample (Volgenau et al. 2006).

To search for binary protostars and derive kinematic properties of
these systems, we have started a program to observe, at high
angular resolution, a number of isolated low-mass prestellar and
protostellar molecular cloud cores, conducted at the Owens Valley
Radio Observatory (OVRO) millimeter array (this work; hereafter
called Paper I) and now continued with ATCA (Australia Telescope
Compact Array) and PdBI (IRAM Plateau de Bure Interferometer)
arrays (Paper II\&III, in prep.). In this paper we present
N$_{2}$H$^{+}$ results for 9 protostellar cores observed with
OVRO. In section 2 we describe the target list, observations, and
data reduction. Observational results are presented in section 3.
We give a detailed description of individual sources and discuss
the implications of our results for binary star formation models
in section 4. The main conclusions of this study are summarized in
section 5.

\section{OBSERVATIONS AND DATA REDUCTION}

For this survey we selected a number of well-isolated and nearby
low-mass protostellar cores. Priority was given to sources which
were already well-characterized either by our own previous
observational data or in the literature. Most sources are Class 0
objects, which represent the youngest protostars at an age of a
few $\times$ 10$^4$ yr. While they may already be too evolved to
represent the true initial conditions, these sources provide an
opportunity to probe the earliest and most active stage of the
star formation process, where most of the initial information is
still preserved. The target list and basic properties of the
sources are summarized in Table\,1.

For studying the gas kinematics and derive rotation curves, we
choose to observe the N$_{2}$H$^{+}$\,(1-0) hyperfine structure
line complex at 93.1378\,GHz. N$_{2}$H$^{+}$ is known to be a
selective tracer of cold, dense, and quiescent gas and is
particularly suitable for studying the structure and kinematics of
cold star-forming cores (Turner \& Thaddeus 1977; Womack, Ziurys,
\& Wychoff 1992; Bachiller 1996; Caselli et al. 2002). It is the
most reliable tracer of the gas kinematics in pre- and
protostellar cores for three reasons: (1) compared to other
molecules, it depletes much later and more slowly onto grains
(Bergin \& Langer 1997), (2) it is formed where CO is depleted and
thus traces only the dense cores and not the wider envelope and is
thus perfectly suited for interferometric observations, and (3)
with its seven hyperfine components within 17 km s$^{-1}$
(including optically thin and moderately optically thick
components; Caselli et al. 1995) it provides much more precise and
reliable kinematic information than a single line, even with a
moderate signal-noise ratio.

Observations were carried out with the OVRO array of six 10.4 m
telescopes during 7 observing seasons between 1999 and 2002. Five
different array configurations (C, L, E, H, and U) were used, with
baselines ranging from 18 to 480 m. All antennas were equipped
with cooled SIS receivers which provided average system
temperatures of $\sim$ 300$-$400 K at the observing frequency. A
digital correlator was centered at 93.1378 GHz. Spectral
resolution and bandwidth were $\sim$ 0.2 km s$^{-1}$ and 25 km
s$^{-1}$, respectively. Amplitude and phase were calibrated
through frequent observations of quasars nearby to each source,
typically every 20 minutes, resulting in an absolute position
uncertainty of $\leq$ 0\farcs2. The flux density scale was
calibrated by observing Neptune and Uranus. The estimated
uncertainty is $<$ 20\%. Observing parameters are summarized
in Table\,2. The thermal dust continuum
emission was measured simultaneously with the N$_{2}$H$^{+}$ and
other line observations using a separate continuum correlator. The
combined continuum results will be published in a separate paper
(Launhardt et al., in prep.).

The raw data were edited and calibrated using the MMA software
package (Scoville et al. 1993), and synthesized images were
produced using MIRIAD (Sault et al. 1995) and its CLEAN algorithm,
with ``robust" weighting of the visibilities (Briggs et al. 1999).
The cleaned and restored maps have effective synthesized beam
sizes of 4$-$8$''$ and 1$\sigma$ rms levels of $\sim$ 70 mJy/beam
(see Table~2). Further analysis and figures were done with the
GILDAS\footnote{http://www.iram.fr/IRAMFR/GILDAS} software
package. With Class (part of GILDAS), we have developed a
semi-automatic fitting routine which allows the derivation of
reliable and very accurate velocity fields from the 7-hyperfine
component line complex. Assuming that bulk rotation is the
dominating motion, the velocity fields are then used to derive the
rotation axis and the specific angular momentum of the cores.

\section{RESULTS}
\subsection{Morphology of N$_2$H$^+$ Cores}
N$_2$H$^+$ emission is detected from all nine targeted objects.
Figure\,1 shows the distribution of the velocity-integrated
intensity of N$_2$H$^+$ towards the nine cores. The emission was
integrated over all seven components, using frequency masks that
completely cover velocity gradients within the sources. All
objects are spatially associated with mm dust continuum sources,
indicative of embedded protostars and their accretion disks. The
positions of the mm continuum sources are indicated by crosses in
Figure\,1 (Launhardt et al. 2007, in prep.). We find that in seven
of the nine objects the mm continuum source lies within the half
maximum level of the N$_2$H$^{+}$ emission. The two exceptions are
CB\,188 (the only Class I object in our sample) and CB\,244. This
good general agreement indicates that N$_2$H$^{+}$ cannot be
significantly depleted like, e.g., CO and CS (see also Bergin et
al. 2001; Caselli et al. 1999).

We also measured mean FWHM radii of the integrated N$_{2}$H$^{+}$
emission. The mean FWHM core radii $R$ were measured as
$A^{1/2}/\pi$, where $A$ is the core area at the half maximum
level, corrected for the beam size. Except for CB\,188, all
sources exhibit quite compact N$_2$H$^+$ emission regions with
mean FWHM radii of 1200$-$3500\,AU (see Table 4). The average
radius in our sample is $\langle$$R$$\rangle$ = 2000 $\pm$ 800 AU,
which is much lower than the average value found by Caselli et al.
(2002; hereafter CBMT02) for their sample of starless cores
($\sim$ 10000 AU) observed with single-dish observations (beam
size $\sim$ 54$''$), but similar to the radius of Class 0 sources
IRAM\,04191+1522 ($\sim$ 2400 AU) and NGC\,1333 IRAS\,4B ($\sim$
1800 AU) observed with PdBI by Belloche et al. (2002) and Di
Francesco et al. (2001), respectively.

However, when viewed in detail, all sources in our sample show a
complex, often multi-peaked structure. For example, IRAS\,03282,
IRAS\,04166, and CB\,230 each have two separated peaks in the
region enclosed by the half maximum intensity level, with the mm
continuum source located between the two peaks. In RNO\,43,
CB\,68, CB\,224, and CB\,244 the main peak of N$_2$H$^{+}$
emission offsets by 5$-$10$''$ from the mm continuum sources. Only
in L723\,VLA2 there is no positional discrepancy between the
N$_2$H$^{+}$ and mm continuum emission. The individual sources are
discussed in detail in Section 4.1.

\subsection{Masses and Column Densities}

Figure 2 shows the N$_2$H$^{+}$ spectra at the position of maximum
intensity in each map. The N$_{2}$H$^{+}$ (1$-$0) line complex
consists of 7 hyperfine structure components, which have been
detected in all sources. However, in several sources like, e.g.,
L723\,VLA2 (see below), the line width is larger than the
separation between hyperfine components, so that some lines are
blended. The hyperfine fitting program in CLASS (Forveille et al.
1989), with the frequencies adopted from Caselli et al. (1995) and
weights adopted from Womack et al. (1992), has been used to
determine LSR velocities ($V_{LSR}$), intrinsic line width
($\triangle$$v$; corrected for instrumental effects), total
optical depths ($\tau_{tot}$), and excitation temperatures
($T_{ex}$). These parameters are listed in Table\,3. Here
$\tau_{tot}$ is the sum of the peak optical depths of the seven
hyperfine components (see Benson \& Myers 1989). The optical depth
of the main N$_2$H$^+$ ($J\,F_1\,F = 1\,2\,3 \longrightarrow
0\,1\,2$) component, which is equal to 0.259$\tau_{tot}$, is found
to be small ($\leq$ 0.5) at the intensity peak for all sources.
Hence the N$_2$H$^{+}$ emission can be considered optically thin
everywhere. The excitation temperature, $T_{ex}$, was calculated
to be 4.1$-$5.7 K at the peak positions, using a main-beam
efficiency $\eta_{B}$ = 0.7 (Padin et al. 1991). The average
$\langle T_{ex} \rangle$ $\sim$ 4.9 K in our sample is similar to
what has been found with single-dish observations for dense cores
by CBMT02 ($\sim$ 5.0 K).

Assuming that the observed N$_{2}$H$^{+}$ line widths are not
dominated by systematic gas motions, the virial mass of the cores
has been calculated as:

\begin{equation}
M_{vir} = \frac{5}{8ln2}\frac{R \triangle v_{m}^2}{\alpha_{vir}
G},
\end{equation}
where $G$ is the gravitational constant, $R$ is the FWHM core
radius, and $\triangle v_{m}$ is the line width of the emission
from an ``average" particle with mass $m_{ave}$ = 2.33 amu
(assuming gas with 90\% H$_2$ and 10\% He). The coefficient
$\alpha_{vir}$ = (3 $-$ $p$)/(5 $-$ 2$p$), where $p$ is the
power-law index of the density profile, is corrected for
deviations from constant density (see Williams et al. 1994). In
our calculations, we assume $p$ = 1.5 and $\alpha_{vir}$ = 0.75
(see Andr\'{e}, Ward-Thompson, \& Barsony 2000). $\triangle v_{m}$
can be derived from the observed spectra by

\begin{equation}
\triangle v_{m}^2 = \triangle v_{obs}^2 +
8ln2\frac{kT_{ex}}{m_H}(\frac{1}{m_{av}}-\frac{1}{m_{obs}}),
\end{equation}
where $\triangle v_{obs}$ is the observed mean line width
(obtained through Gaussian fitting to the distribution of line
widths vs. solid angle area; see Table~3 and $\S$3.4) and $m_{obs}$
is the mass of the emitting molecule (here we use $m_{N_2H^+}$ =
29 amu). The corresponding hydrogen density of the $n_{H_2}$ has
been calculated assuming a uniform density spherical core with
radius $R$ (given in Table\,4). The derived virial masses
$M_{vir}$ in our sample range from 0.3 to 1.2\,$M_\odot$, with a
mean value of 0.6\,$M_\odot$. The corresponding hydrogen densities
$n_{H_2}$ range from 7.4 to 81 $\times$ 10$^{5}$ cm$^{-3}$. Both
$M_{vir}$ and $n_{H_2}$ values for each source are listed in
Table\,4.

The N$_2$H$^+$ column density has been calculated independently
from the line intensity using the equation given by Benson et al.
(1998):

\begin{equation}
N(N_{2}H^{+}) = 3.3 \times 10^{11} \frac{{\tau}{\bigtriangleup}v
T_{ex}}{1-e^{-4.47/T_{ex}}}\,(cm^{-2}),
\end{equation}
where $\tau$ is the total optical depth, $\triangle v$ is the
intrinsic line width in km s$^{-1}$, and $T_{ex}$ is the
excitation temperature in K. The gas-phase N$_{2}$H$^{+}$ mass of
the core can then be calculated from
$M_{N_{2}H^{+}}~\approx~N(N_2H^+)_{peak}\,\times\,m_{N_2H^+}\,\times\,d^{2}\,\times\,\Omega$$_{FWHM}$,
where $d$ is the distance from the sun and $\Omega_{FWHM}$ is the
area enclosed by the contour level at 50\% of the peak value for
each core.

From the ratio of N$_{2}$H$^{+}$ mass to virial mass, we derived
the average fractional abundance of N$_{2}$H$^{+}$ in each core:
$X(N_{2}H^{+})$ = $M_{N_{2}H^{+}}$/$M_{H_2}$, where $M_{H_2}$ =
$M_{vir}$/1.36, the factor 1.36 accounting for He and heavier
elements. The values are listed in Table\,4. The average value
$\langle$$X(N_{2}H^{+})$$\rangle$ $\sim$ 3.3 $\times$ 10$^{-10}$
in our sample is close to that found by CBMT02 for their sample of
dense cores ($\sim$ 3 $\times$ 10$^{-10}$).


\subsection{Velocity Fields}

Based on the hyperfine fitting program in Class, we have developed
a semi-automated quality control and spectra fitting routine that
computes the mean radial velocity, line width, and line intensity
at each point of the map where N$_{2}$H$^{+}$ is detected ($>$
2$\sigma$ noise). Figure 3 shows the mean velocity fields for the
8 Class 0 objects in our sample, obtained from this line fitting
routine. The N$_{2}$H$^{+}$ emission from CB\,188, the only
evolved (Class I) object in our sample, is too faint and fuzzy to
give a reliable velocity field.

In Class\,0 protostars, the effects of infall, outflow,
rotation, and turbulence are generally superimposed. Of these,
turbulence and infall normally broaden the lines but do not produce
systematic velocity gradients. On the other hand, systematic
velocity gradients are usually dominated by either rotation or outflow.
In Fig.\,3, we therefore also show the outflow information for each source.
We want to mention that many other studies usually do not take this kind
of caution. Figure 3 shows that 5 objects
(IRAS\,04166, RNO\,43, L723\,VLA2, CB\,230, and CB\,224) have
well-ordered velocity fields with symmetrical gradients, while 3
objects (IRAS\,03282, CB\,68, and CB\,244) have more complex
velocity fields. Two objects (RNO\,43 and CB\,230) with
well-ordered velocity fields have gradients roughly perpendicular
to the axis of outflow, while 2 objects (IRAS\,04166 and
L732\,VLA2) have gradients basically parallel and one object
(CB\,224) anti-parallel to the outflow direction.

Assuming the mean direction of the bulk angular momentum
is preserved in the collapse from the core to the disk, we would
expect outflows to emerge perpendicular to the rotation velocity gradients.
Thus, if the velocity field is dominated by rotation,
we would see the velocity gradient to be perpendicular to the outflow axis, 
like seen in RNO\,43 and CB\,230. If gradients are parallel to
the outflow axis and with the same orientation, 
it is likely that we see outflow motions rather than rotation,
like in IRAS\,04166 and L723\,VLA2. In these cases, 
a possible underlying rotation velocity gradient must 
be smaller than the observed effective gradient. 
Following these arguments, we treat those gradients parallel 
to the outflow as upper limits to the rotation gradient.  
The details of the velocity field for each source are described in $\S$4.1.

A least-squares fitting of velocity gradients has been performed
for the objects in our sample using the routine described in
Goodman et al. (1993; hereafter GBFM93). The fitting results are
summarized in Table\,5. Listed are in column (2) the mean velocity
of the cores, in columns (3) and (4) the magnitude of the velocity
gradient $g$ and its direction $\Theta_{g}$ (the direction of
increasing velocity, measured east of north), and in column (5)
the total velocity shift across the core $gr$ (i.e., the product
between $g$ and core size $R$).

\subsection{Line Widths}

Figure 4 shows the distribution of line widths vs. solid
angle area in the maps for the 9 protostars in our sample. The
mean line width for each source is then derived through Gaussian
fitting to the distribution and is listed in Table~3. We find that
the mean line width in our sample is 0.29 $-$ 0.51 km s$^{-1}$,
with an average of $\sim$ 0.37 km s$^{-1}$.

The FWHM thermal line width for a gas in LTE at kinetic
temperature $T_{K}$ is given by
\begin{equation}
\triangle v_{th}^2 = 8ln2\frac{kT_{K}}{m_{obs}},
\end{equation}
where $k$ is the Boltzmann constant and $m_{obs}$ is the mass of
the observed molecule. Assuming a kinetic gas temperature of 10\,K
(see Benson \& Myers 1989), the thermal contribution to the
N$_{2}$H$^{+}$ line width is $\sim$ 0.13 km s$^{-1}$. The typical
non-thermal contribution to the line width ($\triangle v_{NT} =
\sqrt{\triangle v_{obs}^2 - \triangle v_{th}^2}$) is then $\sim$
0.35 km s$^{-1}$ , which is about 2.5 times larger than the
thermal line width (see discussion in $\S$4.2).


Figure 5 shows the spatial distribution of the N$_{2}$H$^{+}$ line
width for those 6 Class 0 sources which are not highly elongated
(axial ratios $\leq$ 2). In most cases the line widths are roughly
constant within the interiors of the cores and broader line widths
occur only at the edges. This is consistent with both the NH$_{3}$
observations by Barranco \& Goodman (1998) and the physical
picture described by Goodman et al. (1998) that the star-forming
dense cores are ``velocity coherent" regions of nearly constant
line width. The exception, CB\,230, is discussed in $\S$4.1.8


\section{DISCUSSION}
\subsection{Description of Individual Sources}
\subsubsection{IRAS 03282+3035}

IRAS\,03282 is one of the youngest known Class 0 protostars
(Andr\'{e} et al. 2000). It is located in the western part of the
Perseus molecular cloud complex at a distance of $\sim$ 300 pc
(Motte \& Andr\'{e} 2001). Its bolometric luminosity and total
envelope mass are estimated to be $L_{bol}$ $\sim$
1.2\,$L{_\odot}$ and $M_{env}$ $\sim$ 2.9\,$M{_\odot}$,
respectively (Shirley et al. 2000). IRAS\,03282 drives a highly
collimated bipolar molecular outflow at P.A. $\sim$
$-$37$^{\circ}$ with kinematic age $\sim$ 10$^{4}$ yr (Bachiller
et al. 1994).

The N$_{2}$H$^{+}$ intensity map of IRAS\,03282 shows a molecular
cloud core that is hour-glass shaped and elongated along P.A.
$\sim$ 30 degree, i.e., nearly perpendicular to the axis of the
large-scale CO outflow (see Fig.\,3). We interpret the hourglass
shape as effect of the outflow, which re-releases CO from grain
surface back into the gas phase. CO, in turn, destroys the
N$_{2}$H$^{+}$ molecule (Aikawa et al. 2001). Our OVRO 1.3 mm dust
continuum images reveal two mm sources located between the two
N$_{2}$H$^{+}$ emission peaks (see Fig.\,3). The mass ratio and
the angular separation of this binary protostar are 0.23 and
1\farcs5 ($\sim$ 450 AU at a distance of 300 pc), respectively
(Launhardt et al. 2007, in prep.; hereafter LSZ07). The velocity
field of IRAS\,03282 does not show a symmetrical gradient and the
red-shifted N$_{2}$H$^{+}$ emission lobe to the northwest of the
mm sources matches exactly the red-shifted $^{13}$CO lobe,
implying that the N$_{2}$H$^{+}$ emission in this region is
affected by the outflow. Correspondingly, we treat the observed
total velocity gradient as upper limit to rotation (see $\S$3.3).
Figure 5 shows the line width distribution of IRAS\,03282. The
average line width across the source is $\sim$ 0.4 $\pm$ 0.1 km
s$^{-1}$ and reaches 0.7 km s$^{-1}$ towards the southern edge of
the core.

\subsubsection{IRAS 04166+2706}

IRAS\,04166 is a Class 0 protostar associated with the small dark
cloud B\,213 in the Taurus molecular cloud complex at a distance
$\sim$ 140 pc (Mardones et al. 1997). Its bolometric luminosity
and total envelope mass are $L_{bol}$ $\sim$ 0.4\,$L{_\odot}$ and
$M_{env}$ $\sim$ 1.0\,$M{_\odot}$, respectively (Shirley et al.
2000). IRAS\,04166 drives a highly collimated, extremely high
velocity (up to 50 km s$^{-1}$) bipolar molecular outflow at P.A.
$\sim$ 30$^{\circ}$ (Tafalla et al. 2004).

The N$_{2}$H$^{+}$ intensity map of IRAS\,04166 shows the same
hourglass shape and orientation relative to the outflow as
IRAS\,03282. We interpret it in the same way. The N$_{2}$H$^{+}$
velocity field of IRAS\,04166 shows a symmetric gradient, which
however matches the red- and blue-shifted CO outflow lobes (see
Fig.~3). As for IRAS\,03282, we treat the observed total velocity
gradient as upper limit to rotation.

\subsubsection{RNO\,43\,MM}

The millimeter continuum source RNO\,43\,MM is associated with the
dark cloud L\,1582B and the very cold IRAS point source 05295+1247
(detected only at 60 and 100 $\mu$m). The cloud is physically
connected to the $\lambda$\,Ori molecular ring located at a
distance of $\sim$ 400 pc (Zinnecker et al. 1992). RNO\,43\,MM is
the origin of a 3.4 pc long Herbig-Haro flow (HH\,243, HH\,244,
and HH\,245; Reipurth et al. 1997) and drives a collimated bipolar
molecular outflow at P.A. $\sim$ 60$^{\circ}$ (Arce \& Sargent
2004). From its spectral energy distribution (SED) and the ratio
of sub-mm to bolometric luminosity, RNO\,43MM has been classified
a Class 0 source (Bachiller 1996). Its bolometric luminosity and
total envelope mass are estimated to be $L_{bol}$ $\sim$
6.0\,$L{_\odot}$ and $M_{env}$ $\sim$ 0.4\,$M{_\odot}$,
respectively (Zinnecker et al. 1992).

Our N$_{2}$H$^{+}$ intensity map of RNO\,43MM shows a V-shaped
core with two lobes, extending $\sim$ 20$''$ to the north and
west, respectively (see Fig.\,1). The western lobe exhibits much
higher (redder) mean velocities than the rest of the core, and is
outside the plot scale of Fig.~3 (11 up to 20 km s$^{-1}$). This
jump in velocity probably means that the western lobe belongs to
another molecular cloud layer in this direction. In the inner core
region (included at 50\% contour level), the velocity field
exhibits a symmetric structure with a gradient of $\sim$ 5.8 km
s$^{-1}$ pc$^{-1}$ at P.A. $\sim$ $-$13$^{\circ}$, approximately
perpendicular to the axis of outflow, suggesting rotation (see
Fig.\,3). The total velocity shift across the inner core is $\sim$
0.4 km s$^{-1}$.

\subsubsection{CB 68}

CB\,68 (LDN 146) is a small Bok globule located in the outskirts
of the $\rho$ Oph dark cloud complex at a distance of 160 pc
(Clemens \& Barvainis 1988; Launhardt \& Henning 1997). The dense
core of the globule exhibits strong, extended, centrally peaked
sub-mm/mm dust continuum emission (Launhardt \& Henning 1997;
Launhardt et al. 1998; Vall\'{e}e, Bastien, \& Greaves 2000) and
is associated with the cold IRAS point source 16544--1604. The
central source, which was classified as a Class 0 protostar,
drives a weak, but strongly collimated bipolar molecular outflow
at P.A. $\sim$ 142$^{\circ}$ (Wu et al. 1996; Mardones et al.
1997; Vall\'{e}e et al. 2000).

Our N$_{2}$H$^{+}$ intensity map shows a compact source of about
9$''$ FWHM radius, which peaks very close to the mm continuum
position. In addition, it exhibits two armlike extensions to
northeast and southwest. The velocity field of CB\,68 is
relatively complicated and shows no systematic gradient which
could be interpreted as rotation. There is also no clear
correlation with the outflow. As for IRAS\,03282, we derive only
an upper limit for the rotation velocity gradient.

\subsubsection{L723 VLA2}

L723 is a small isolated dark cloud located at a distance of
300$\pm$150 pc (Goldsmith et al. 1984). The IRAS point source at
the center of the cloud core is associated with strong
far-infrared (Davidson 1987), sub-mm (Shirley et al. 2000) and mm
continuum emission (Cabrit \& Andr\'{e} 1991; Reipurth et al.
1993) and was classified as Class 0 protostar with a bolometric
luminosity $L_{bol}$ $\sim$ 3.0\,$L{_\odot}$ and a circumstellar
mass $M_{env}$ $\sim$ 1.2 $M_{\odot}$ (LSZ07). Anglada et al.
(1991) detected two radio continuum sources (VLA1 and VLA2) with
15$''$ separation, both located within the error ellipse of the
IRAS position. However, only VLA2 was found to be associated with
dense gas in the cloud core (Girart et al. 1997). Centered at the
IRAS position is a large quadrupolar molecular outflow with two
well-separated pairs of red and blue lobes: a larger pair at P.A.
$\sim$ 110$^{\circ}$ and a smaller pair at P.A. $\sim$
30$^{\circ}$ (Lee et al. 2002 and references therein). While
several scenarios have been proposed to explain the quadrupolar
morphology of the large-scale CO outflow, the discovery of a
thermal radio jet at the position of VLA2 (Anglada et al. 1996)
and of a large-scale H$_2$S(1) bipolar flow (Palacios \& Eiroa
1999), both aligned with the larger pair of CO outflow lobes at
P.A. $\sim$ 110$^{\circ}$, as well as new high-resolution CO
observations (Lee et al. 2002) clearly favor the presence of two
independent flows. Indeed, our OVRO mm continuum observations
reveal two compact sources in L723 VLA2 separated by $\sim$
3\farcs2 (960 AU at a distance of 300 pc), supporting strongly the
scenario that the quadrupolar outflow is driven by a binary
protostar system (see Launhardt 2004).

The N$_{2}$H$^{+}$ intensity map of L723 VLA2 shows a compact
source of about 8$''$ FWHM diameter, which peaks between the two
mm continuum sources. In addition, it shows a long extension to
the northwest along the direction of larger outflow. The mean
velocity difference between the two continuum positions is $\sim$
0.4 km s$^{-1}$. However, the overall velocity field of L723 VLA2
shows that the red-shifted N$_{2}$H$^{+}$ emission matches exactly
the red-shifted emission of two outflows and the velocity gradient
is basically in the same direction as the outflow, suggesting that
the outflow has a strong effect on the N$_{2}$H$^{+}$ emission. As
for IRAS\,03282, we therefore treat the observed total velocity
gradient as upper limit to rotation. The line widths distribution
of L723 VLA2 is shown in Fig.~5. The line widths are about 0.5 km
s$^{-1}$ in the northwest extension, 0.6$-$1.0 km s$^{-1}$ across
the core, and reach 1.5 km s$^{-1}$ towards the eastern edge of
the core. We suggest that the relatively large line widths in this
source are the result of strong outflow-envelope interaction.

\subsubsection{CB 188}

CB\,188 is a small dark globule at a distance of $\sim$ 300 pc,
which harbors a Class I YSO (Launhardt \& Henning 1997). A small
($\leq$ 2$'$) CO outflow with overlapping red and blue lobes was
detected by Yun \& Clemens (1994), suggesting the YSO is seen
close to pole-on. In our N$_{2}$H$^{+}$ intensity map, CB\,188
shows a complex clumpy structure (see Fig.\,1). It also has the
weakest N$_2$H$^{+}$ emission in our survey (also see the spectra
in Fig.\,2). This could imply that N$_2$H$^{+}$ is destroyed
during the evolution from Class 0 to Class I, e.g., due to release
of CO from grain surfaces in outflows (see Aikawa et al. 2001). We
do not further discuss this source in this paper.

\subsubsection{CB 224}

CB\,224 is a Bok globule located at a distance of $\sim$ 400 pc
(Launhardt \& Henning 1997). Two mm sources were detected in our
OVRO survey at an angular separation of 20$''$ (LSZ07). The
northeast mm source (not shown here) is associated with a cold
IRAS source 20355+6343 (3.9\,$L_\odot$; Launhardt \& Henning
1997), but has no N$_{2}$H$^{+}$ emission detected. The southwest
source shown in our images, which is classified a Class 0 object
(LSZ07), drives a collimated $^{13}$CO bipolar outflow at P.A.
$\sim$ $-$120$^{\circ}$ (Chen et al. 2007, in prep.).

The N$_{2}$H$^{+}$ intensity map of CB\,224 shows the same
hourglass morphology perpendicular to the outflow as IRAS\,03282
and IRAS\,04166 (but slightly more asystematic in intensity).
There is a clear and systematic velocity gradient across the core,
but the lines of constant velocity are curved in a ``C" shape (see
Fig.\,3). The velocity field cannot be solely interpreted by
rotation. The CO outflow seems to have no effect on the
N$_{2}$H$^{+}$ velocity field because it is oriented in the
opposite direction. We speculate that the velocity field is due to
a combination of rotation and core contraction. The HCO$^{+}$
observations in De Vries et al. (2002) indeed indicated signs of
infall motions in CB\,224.

\subsubsection{CB 230}

CB\,230 (L 1177) is a small, bright-rimmed Bok globule at a
distance of 400$\pm$100 pc (Wolf et al. 2003). The globule
contains a protostellar core of total mass $\sim$ 5$M{_\odot}$ and
exhibits signatures of mass infall (Launhardt et al. 1997, 1998,
2001). Magnetic field strength and projected direction in the
dense core are $B$ = 218 $\pm$ 50 mG and P.A. = $-$67$^{\circ}$,
respectively (Wolf et al. 2003). The dense core is associated with
a large-scale collimated CO outflow at P.A. = 7$^{\circ}$ of
dynamical age $\sim$ 2 $\times$ 10$^4$ yr (Yun \& Clemens 1994;
Chen et al. 2007, in prep.). The Mid-IR image of CB\,230 suggests
the presence of two deeply embedded YSOs separated by $\sim$10$''$
(Launhardt 2004). Only the western source was detected at 1.3\,mm
and 3\,mm dust continuum, suggesting that the mass of a possible
accretion disk around eastern source is below the detection limit.
Two bright near-infrared reflection nebulae are associated with
the embedded YSOs, but the stars are not directly detected at
wavelengths shorter than 5 $\mu$m. CB\,230 is probably a
transition object between Class 0 and I.

Our N$_{2}$H$^{+}$ intensity map shows that the molecular cloud
core is elongated East-West. The velocity field map shows a clear
velocity gradient across the core of $\sim$ 8.8 km s$^{-1}$
pc$^{-1}$ increasing from east to west along the long axis, i.e.,
roughly perpendicular to the outflow axis. This strongly supports
the view that the two MIR sources form a protobinary system which
is embedded in the N$_{2}$H$^{+}$ core. This core in turn rotates
about an axis perpendicular to the connecting line between the two
protostars and parallel to the main outflow. Fig.~5 shows that the
line width distribution exhibits a strong peak at the position of
the 3\,mm continuum source. Together with the velocity field shown
in Fig.\,3, this can be understood as the result of Keplerian
rotation. Morphology and kinematics of this source will be
discussed more detailed in another paper.

\subsubsection{CB 244}
CB\,244 (L\,1262) is a Bok globule located at a distance of $\sim$
180\,pc (Launhardt \& Henning 1997). It is associated with a faint
NIR reflection nebula and a cold IRAS point source, and was
classified as Class 0 protostar. Its bolometric luminosity and
total envelope mass are estimated to be $L_{bol}$ $\sim$
1.1\,$L{_\odot}$ and $M_{env}$ $\sim$ 3.3\,$M{_\odot}$,
respectively (Launhardt \& Henning 1997). CB\,244 drives a bipolar
molecular outflow at P.A. $\sim$ $-$130$^{\circ}$ (Yun \& Clemens
1994; Chen et al. 2007, in prep.).

The N$_{2}$H$^{+}$ intensity map of CB\,244 shows an elongated
structure in the direction from Northwest to Southeast,
approximately perpendicular to the direction of outflow (see
Fig.\,3). The velocity field of CB\,244 exhibits a complicated
structure and maybe dominated by effects other than rotation. As
for other sources, we assume that the total observed velocity
gradient puts an upper limit to the rotation.
The distribution of line widths for this source shows
two distinct peaks (see Fig.\,4). The smaller line widths originate
from the southeastern part of the core where the protostar is embedded,
while the larger widths are found in the northwestern extension only
(see Fig.\,1). Therefore, we adopted the smaller peak value as
representative for CB\,244.

\subsection{Turbulent Motions}

At a kinetic gas temperature of 10\,K, the typical non-thermal
line width in our sample (0.35 km s$^{-1}$) is about 2.5 times
lager than the thermal line width (0.13 km s$^{-1}$) (see
$\S$3.4). The origin of the non-thermal line width in such cores
is subject of an ongoing debate in the literature. Generally,
turbulence is suggested to be the main contribution (see e.g.,
Barranco \& Goodman 1998 and Goodman et al. 1998), but infall,
outflow, and rotation may also contribute to the non-thermal line
width.

With the exception of L723 VLA2, there appears to be no spatial
correlation between regions of increased line width and outflow
features (see Fig.\,5). CB\,230 may represent a special case where
Keplerian rotation of a large circumstellar disk causes the large
non-thermal line width in the central region (see Fig.\,5). To
avoid a bias from localized line-broadening due to outflows and/or
Keplerian rotation, we estimated the observed mean line width
through Gaussian fitting to the distribution of line widths vs.
area in the maps (see Fig.\,4). The non-thermal
contribution to these mean line widths are listed in Table~3. We
assume that these mean non-thermal line widths are dominated by
turbulence. The thermal FWHM line width of an ``average" particle
of mass 2.33\,m$_{H}$, which represents the local sound speed,
would be $\sim$ 0.44 km s$^{-1}$ at 10\,K. The mean observed
non-thermal line width is $\sim$ 1.3 times smaller than this
value, which means that turbulence in these cores is subsonic.

Figure~6 shows the distribution of non-thermal line width
$\triangle$$v_{NT}$ with core size $R$ for the objects in our
sample, together with the dense cores from GBFM93 and CBMT02. It shows
that high-level (supersonic) turbulence normally occurs in large-scale cores only. 
In the cores traced by NH$_{3}$ ($R$ $>$ 20000\,AU
$\sim$ 0.1\,pc), the non-thermal line widths decrease with core size
with a power-low index $\sim$ 0.2, while in those traced by
N$_{2}$H$^{+}$ (CBMT02; 5000\,AU $<$ $R$ $<$
20000\,AU), the line widths decrease with an index of $\sim$
0.5. This suggests that non-thermal motions are more quickly
damped in smaller cores (see also Fuller \& Myers 1992). 
Comparing our data with CBMT02, 
we find that the relation between line width and core size
no longer holds at $R$ $<$ 10000\,AU, 
suggesting the inner parts of dense cores are ``velocity coherent".
The radius at which the gas becomes ``coherent" is less than 0.1\,pc, 
as suggested in Goodman et al. (1998). 
Also note that the mean non-thermal line width in our 
sample (0.35 km s$^{-1}$) is even larger than the widths in some 
larger scale dense cores (see Fig.\,6). We speculate that the heating from an
internal protostar, as well as related activities, like e.g., infall and/or
outflow, contribute to the line widths in these protostellar cores.

\subsection{Systematic Gas Motions: Fast Rotation of Protostellar Cores}

For the fragmentation model of binary star formation, some initial
angular momentum must be present; otherwise the cores will
collapse onto a single star. The source of this angular momentum
is generally suggested to be the bulk rotation of the core. As
suggested in the earlier study by GBFM93, rotation is a common
feature of dense cores in molecular clouds. In our observations,
most objects show well-ordered velocity fields with symmetric
gradients and the motions of several objects could be interpreted
as bulk rotation. The preliminary results from our ATCA and PdBI
observations show similar well-ordered velocity fields (Paper
II\&III, in prep.), supporting the view in GBFM93.

Up to now only a few Class 0 objects have been studied in detail
kinematically. Typical examples are the nearby, isolated object
IRAM\,04191+1522 (IRAM\,04191 for short; see Belloche et al.
2002), or NGC\,1333\,IRAS\,4A (IRAS\,4A for short; see Di
Francesco et al. 2001 and Belloche et al. 2006), which is actually
a binary protostar at a separation of $\sim$ 600 AU (Looney et al.
2000). Below, we analyze our 8 Class 0 objects together with these two
sources. Of the 8 Class 0 targets, CB\,230, IRAS\,03828, and
L723\,VLA2 are resolved as binary protostars by our mm
observations, while IRAS\,04166, RNO\,43, CB\,68, CB\,224, and
CB\,244 remain single or unresolved.
We treat RNO\,43 and CB\,230 as objects which provide
reliable rotation velocity gradients and take the other 6 measurements as
upper limits. The ASURV\footnote{ASURV Rev.1.2 (LaValley, Isobe \& Feigelson 1992)
is a software package which implements the methods presented 
in Feigelson \& Nelson (1985). For details see http://astrostatistics.psu.edu/.} 
software package was used for the statistical
analysis of the results. This package performs a ``survival analysis''
which takes into account upper limits and allows to compute a statistical
sample mean. It must, however, be noted that two real
measurements and 6 upper limits are not sufficient to derive statistically
significant correlations.

The velocity gradients derived in our sample range from 5.8 to
$\leq$\,24.2 km s$^{-1}$ pc$^{-1}$ (see Table~5). 
The mean value derived by the cumulative Kaplan-Meier (KM) 
estimator in ASURV is 7.0$\pm$0.8 km s$^{-1}$ pc$^{-1}$.
This result is consistent with the gradients derived in IRAM\,04191 ($\sim$ 7 km
s$^{-1}$ pc$^{-1}$) and IRAS\,4A ($\sim$ 9.3 km s$^{-1}$
pc$^{-1}$), but it is much larger than the velocity gradients of dense
cores derived from single-dish observations in GBFM93 and CBMT02
(1$-$2 km s$^{-1}$ pc$^{-1}$). The correlation between velocity
gradient ($g$) and core size ($R$) is shown in Figure~7. It shows
a clear trend with smaller cores ($R$ $<$ 5000 AU) having larger
velocity gradients. Taking into account only RNO\,43 and CB\,230,
the entire dataset could be fitted by a relation of $g$ $\propto$ $R$$^{-0.6\pm0.1}$, 
which is steeper than the slope of $\propto R^{-0.4}$ obtained by GBFM93 
for the larger scale cores only.
As expected, smaller (more evolved) protostellar cores 
rotate much faster than larger (prestellar) cores.

\subsection{Constraints on Angular Momentum}

Assuming the velocity gradients are due to core rotation, the
specific angular momentum $J/M$ of the objects can be calculated
with the following equation (see GBFM93):

\begin{equation}
J/M = \alpha_{rot}\omega R^2 =
\frac{2}{5+2\alpha}\frac{g}{sini}R^2 \approx \frac{1}{4}gR^2,
\end{equation}
where the coefficient $\alpha_{rot}$ = $2/(5+2p)$, $p$ is the
power-law index of the radial density profile (here we adopt $p$ =
1.5), $g$ is the velocity gradient, and $i$ is the inclination
angle to the line of sight direction. Here we assume $sin\,i$ to
be 1 for all sources. The derived specific angular momenta $J/M$
for our 8 Class 0 protostars are listed in column (6) in Table 5.
We derived values between $<$\,0.12 and 0.45 $\times$ 10$^{-3}$ km 
s$^{-1}$ pc, with a KM sample mean of 0.21$\pm$0.1 $\times$ 10$^{-3}$ km 
s$^{-1}$ pc.

In Figure~8 we show the distribution of specific angular momentum
vs. size scale for Class 0 single (unresolved) and binary
protostars (this work), together with molecular cloud cores, PMS
binary stars, and single stars, etc. The data of NH$_{3}$ dense
cores and N$_{2}$H$^{+}$ starless cores are from GBFM93 and
CBMT02, respectively. For PMS binary stars, the specific orbital
angular momenta are derived as $J/M$ = $\sqrt{GM_{B}D}$$\times
q/(1+q)^2$ (where $M_{B}$ is the total stellar mass, $D$ is the
separation, and $q$ is the mass ratio). Data of Class I and T
Tauri binaries are from Chen et al. (Paper IV, in
prep.)\footnote{The masses of Class I and T Tauri binaries are
dynamic masses. The angular momentum plotted in Fig.\,8 for these
binaries does not include the angular momentum from the stellar
rotation.}; data of very low-mass ($<$ 0.1 $M_\odot$) and brown
dwarf binaries are from Burgasser et al. (2007). The specific
angular momentum of Class I single stars is derived from $J/M$ =
$vR$, where $v$ of 38 km s$^{-1}$ and $R$ of 2.7 $R_\odot$
are mean values from the sample of Class\,I stars
in Covey et al. (2005) and adopted as
representative values here.

As shown in Fig.\,8, there is a strong correlation between $J/M$
and size scale in dense cores ($\geq$ 5000 AU). The data can be
fitted with a power-law correlation of $J/M \sim R^{1.7\pm0.1}$,
which is consistent with the index $1.6\pm0.2$ obtained by GBFM93.
This means that in these dense cores the angular velocity is
locked at a constant value to the first order, which is generally
explained by the mechanism of magnetic braking (see Basu \&
Mouschovias 1994). In addition, Fig.\,8 shows that the mean $J/M$
value in this region is considerably ($\sim$ 2 magnitudes) larger
than the typical orbital angular momentum of T Tauri binary
systems. However, stars are not formed from the entire cloud
(large dense cores), but only from the dense inner $R$ $\leq$ 5000
AU part of the cores that undergoes dynamical gravitational
collapse and decouples from the rest of the cloud. Fig.\,8 shows
that such cores have almost the same specific angular momentum as
the widest (few hundred AU) PMS binaries. But, there is a gap in
size scale between the smallest prestellar cores and the widest
binaries. We find that the specific angular momenta of Class 0
protostellar cores are located in this gap, but nearly
indistinguishable from those of wide PMS binary systems, i.e.,
angular momentum is basically maintained (conserved) from the
smallest prestellar cores via protostellar cores to wide PMS
binary systems. In the context that most protostellar cores are
assumed to fragment and form binary stars, this means that most of
the angular momentum contained in the collapse region is
transformed into orbital angular momentum of the resulting stellar
binary system.

\subsection{Energy Balance}

In this section we estimate the contribution of different terms to
the total energy balance of the protostellar cores in our sample.
In Table~6 we summarize the basic equations and the estimated
ratios for the rotational, thermal, and turbulent energy to the
gravitational potential energy. Here we assume a mean kinetic gas
temperature $\sim$ 10\,K for all objects. The masses and radii
used in the equations are virial masses and FWHM radii listed in
Table~4.

The ratios of thermal and turbulent energy to gravitational
potential energy ($\langle \beta_{therm} \rangle$ $\approx$ 0.26
and $\langle \beta_{turb} \rangle$ $\approx$ 0.15) show that in
these protostellar cores both thermal and turbulent contribution
together appear to dominate the support of the cores, but the
thermal contribution $\sim$ 2 times outweighs turbulence. 
The estimated $\beta_{rot}$ values range from $<$\,0.004 to 0.017, 
with a KM sample mean of $\sim$ 0.007$\pm$0.002, which
is much lower than $\beta_{therm}$ and $\beta_{turb}$. This suggests
that rotation is not dominating in the support of the protostellar
cores. When we apply the equilibrium virial theorem 2[$E_{therm} +
E_{turb} + E_{rot}$] + $E_{grav}$ = 0 (in the absence of magnetic
fields), we find that all the Class 0 protostellar cores in our
sample are slightly supercritical (see Table~6).

Although the rotation energy is relatively small, it is thought to play an
important role in the fragmentation process (see reviews by
Bodenheimer et al. 2000 and Tholine 2002). In general, if
$\beta_{rot}$ is very large ($\beta_{rot} \geq 0.25-0.3$), a gas
cloud can be stable against dynamical collapse, potentially
inhibiting fragmentation and star formation. However, if
$\beta_{rot}$ is very small, a cloud will not have enough
rotational energy to experience fragmentation. Boss (1999) has
shown that rotating, magnetized cloud cores fragment when
$\beta_{rot}$ $>$ 0.01 initially\footnote{Machida et al. (2005)
find that the fragmentation does occur in rotating, magnetized
clouds when $\beta_{rot}$ $\geq$ 0.04, considering magnetic fields
suppress fragmentation.}. It should be noted that three sources in
our sample might have a $\beta_{rot}$ $>$ 0.01. Of these, CB\,230
($\beta_{rot}$ $\sim$ 0.017) and L723 VLA2 ($\beta_{rot}$ $<$
0.014) have been resolved as binary protostars. On the other hand,
IRAS\,03282, a resolved binary protostar, has a very low $\beta_{rot}$ ($\sim$ 0.004),
and IRAS\,04166, a single (unresolved) protostar, has $\beta_{rot}$ value close to 0.01.
Thus, our observations can neither confirm nor disprove a relation
between $\beta_{rot}$ and fragmentation. This could be due to low-number statistics
combined with observational uncertainties.
Figure~9 shows the distribution of 
$\beta_{rot}$ with core size $R$ for our sample, together with the 
cores from GBFM93 and CBMT02. It shows that $\beta_{rot}$ is
roughly independent of $R$, as suggested before by GBFM93.

\section{Summary and Conclusions}

We present N$_2$H$^+$ (1-0) observations of 9 isolated low-mass
protostellar cores using the OVRO mm array. The main conclusions
of this work are summarized as follows:

(1) N$_{2}$H$^{+}$ emission is detected in all target objects and
the emission is spatially consistent with the thermal dust
continuum emission. The mean excitation temperature of the
N$_{2}$H$^{+}$ line is $\sim$ 4.9\,K. The mean FWHM core radius is
$\langle$$R$$\rangle$~=~2000\,$\pm$\,800\,AU. The derived virial
masses of the N$_{2}$H$^{+}$ cores in our sample range from 0.3 to
1.2\,$M_\odot$, with a mean value of 0.6\,$M_\odot$. The
corresponding mean hydrogen number densities range from 10$^{6}$
to 10$^{7}$ cm$^{-3}$. The N$_{2}$H$^{+}$ column densities in our
sample range from 0.6 to 2.8 $\times$ 10$^{12}$ cm$^{-2}$, with a
mean value of 1.4 $\times$ 10$^{12}$ cm$^{-2}$. The average
fractional abundances of N$_{2}$H$^{+}$, calculated by relating
the N$_{2}$H$^{+}$ column densities derived from the line strength
to the virial masses, was found to be $\sim$ 3.3 $\times$
10$^{-10}$. This is consistent with the result obtained in other
surveys with single-dish observations.

(2) The observed mean line widths range from 0.29 to 0.51 km
s$^{-1}$, with a mean value of 0.37 km s$^{-1}$. The non-thermal
contribution is about 2.5 times larger than the thermal line
width, suggesting that the protostellar cores in our sample are
not purely thermally supported. We find that line widths are
roughly constant within the interiors of the cores and larger line
widths only occur at the edges of the cores. We conclude that
turbulence is not negligible but subsonic in the protostellar
cores.

(3) We derive the N$_{2}$H$^{+}$ velocity fields of eight Class 0
protostellar cores. CB\,230 and RNO\,43 show symmetrical velocity
gradients that can be explained by bulk rotation. In L723 VLA2,
IRAS\,03282, and IRAS\,04166, outflow-envelope interaction appears
to dominate the velocity fields. CB\,68, CB\,224, and
CB\,244 show complicated velocity fields, which could be affected
by infall or large-scale turbulence. We argue that in these cores
the observed velocity gradients provide an upper limit to any
underlying bulk rotation.

(4) The velocity gradients over the cores range from 6 to 24 km
s$^{-1}$, with a mean value of $\sim$ 7 km s$^{-1}$ pc$^{-1}$.
This is much larger than what has been found in single-dish
observations of prestellar cores, but agrees with recent
interferometric observations of other Class 0 protostellar cores.
Assuming these gradients are due to rotation, the comparison
between gradients and core sizes suggests that smaller (evolved)
protostellar cores rotate much faster than larger dense
(prestellar) cores. The data could be fitted by a relation of $g
\propto R^{-0.6\pm0.1}$.

(5) We find that in terms of specific angular momentum and size
scale Class 0 protostellar cores fill the gap between dense
molecular cloud cores and PMS binary systems. There appears to be
no evolution (decrease) of angular momentum from the smallest
prestellar cores via protostellar cores to wide PMS binary system.
In the context that most protostellar cores are assumed to
fragment and form binary stars, this means that most of the
angular momentum contained in the collapse region is transformed
into orbital angular momentum of the resulting stellar binary
system.

(6) Both thermal and turbulent energy together dominate the
support against gravity, but the thermal contribution is about 2
times larger than turbulence. All protostellar cores in our sample
are found to be slightly virially supercritical.

(7) The ratio $\beta_{rot}$ of rotational energy to gravitational
energy is relatively small in our sample, ranging from 0.004 to
0.02, with a mean value of 0.007. We find that $\beta_{rot}$ values
in our sample show no clear correlation with observed binary
protostars. On the other hand, the three identified binary
protostars are also not distinguished by $\beta_{turb}$ values.
This could be due to low-number statistics combined with 
observational uncertainties.

\acknowledgments

We thank the anonymous referee for many insightful comments and
suggestions. The Owens Valley millimeter-wave array was supported
by NSF grant AST 9981546. Funding from NASA's {\it Origins of
Solar Systems} program (through grant NAG5-9530) is gratefully
acknowledged. Research at Owens Valley on the formation of young
stars and planets was also supported by the {\it Norris Planetary
Origins Project}. We want to thank A. I. Sargent, who was directly
involved in the early stages of this project, for fruitful
discussions. We thank the OVRO staff for technical support during
the observations. We also thank A. Goodman and S. Schnee for
providing the VFIT routine and thank E. Feigelson for providing the 
ASURV code.


\clearpage

\begin{deluxetable}{lccccccccc}
\tabletypesize{\scriptsize}\tablecaption{\footnotesize Basic
properties of target sources\label{tbl-1}} \tablewidth{0pt}
\tablehead{\colhead{Source}&\colhead{Associated}&\colhead{$D$}
&\colhead{$L_{bol}$}&\colhead{$T_{bol}$}&\colhead{$M_{env}$}&\colhead{Outflow}&\colhead{Infall}
&\colhead{Class}&\colhead{Refs.}\\
\colhead{Name}&\colhead{IRAS source}&\colhead{[pc]}
&\colhead{[L$_{\odot}$]}&\colhead{[K]}&\colhead{[M$_{\odot}$]}&\colhead{}&\colhead{}
&\colhead{}&\colhead{}} \startdata
IRAS\,03282  & 03282+3035   & 300 & 1.2  & 23 & 2.9 & y, bipol, coll. & y & 0   & 1,2,3  \\
IRAS\,04166  & 04166+2706   & 140 & 0.4  & 91 & 1.0 & y, bipol, coll. & y & 0   & 1,3,4  \\
RNO\,43 MM   & 05295+1247   & 400 & 6.0  &    & 0.4 & y, bipol, coll. &   & 0   & 5,6    \\
CB\,68       & 16544$-$1604 & 160 & 1.6  & 74 & 3.5 & y, bipol, coll. &   & 0   & 7,8    \\
L723\,VLA2   & 19156+1906   & 300 & 3.3  & 47 & 7.3 & y, quadrupol.   &   & 0   & 3,9    \\
CB\,188      & 19179+1129   & 300 & 2.6  &    & 0.7 & y, bipol        &   & I   & 8,10    \\
CB\,224      & 20355+6343   & 400 & 16   &    & 6.6 & y, bipol        & y & 0/I & 11,12,13\\
CB\,230      & 21169+6804   & 400 & 8.2  &    & 5.1 & y, bipol, coll. & y & 0/I & 10,11,12 \\
CB\,244      & 23238+7401   & 180 & 1.1  & 56 & 3.3 & y, bipol        & y & 0   & 8,10,14  \\

\enddata
\tablenotetext{a}{The IRAS source is not always associated with
the mm source.} \tablecomments{References. --- (1) Mardones et al.
1997; (2) Bachiller et al. 1994; (3) Shirley et al. 2000; (4) Tafalla
et al. 2004; (5) Zinnecker et al. 1992; (6) Arce \& Sargent 2004;
(7) Vall\'{e}e et al. 2000; (8) Launhardt \& Henning 1997; (9) Lee
et al. 2002; (10) Yun \& Clemens 1994; (11) Wolf et al. 2003; (12)
Launhardt et al. 1998; (13) Chen et al. in prep.; (14) Wang et al.
1995}
\end{deluxetable}


\begin{deluxetable}{lcccccc}
\tabletypesize{\scriptsize} \tablecaption{Target list and summary
of observations\label{tbl-2}} \tablewidth{0pt}
\tablehead{\colhead{Source} &\colhead{Other} &\colhead{R.A. \&
Dec. (1950)$^{a}$} &\colhead{Array} &\colhead{UV coverage}&\colhead{HPBW$^{b}$} &\colhead{rms}\\
\colhead{Name} &\colhead{Name} &\colhead{[h\,:\,m\,:\,s,
$^{\circ}:\,':\,''$]} &\colhead{configuration}
&\colhead{[k$\lambda$]}&\colhead{[arcsecs]}
&\colhead{[mJy/beam]}}\startdata

IRAS\,03282  & ...     & 03:28:15.2, +30:35:14  & CLE   &3--36 & 5.5$\times$4.3 & 76\\
IRAS\,04166  & B\,213  & 04:16:37.8, +27:06:29  & CLU   &3--145& 4.8$\times$4.2 & 45\\
RNO\,43 MM   & L\,1582B& 05:29:30.6, +12:47:35  & CLEH  &3--68 & 5.3$\times$4.7 & 52\\
CB\,68       & L\,146  & 16:54:27.2, $-$16:04:44& CLH   &3--62 & 8.4$\times$5.0 & 71\\
L723\,VLA2   & ...     & 19:15:41.8, +19:06:45  & CLE   &3--36 & 5.5$\times$4.5 & 81\\
CB\,188      & ...     & 19:17:54.1, +11:30:02  & CLEU  &3--127& 4.4$\times$4.1 & 55\\
CB\,224      & L\,1100 & 20:35:30.6, +63:42:47  & CL    &3--36 & 5.4$\times$5.1 & 97\\
CB\,230      & L\,1177 & 21:16:53.7, +68:04:55  & CL    &3--36 & 7.1$\times$6.4 & 63\\
CB\,244      & L\,1262 & 23:23:48.5, +74:01:08  & CL    &3--36 & 6.3$\times$5.2 & 86\\

\enddata
\tablenotetext{a}{Reference position for figures and tables in the
paper.} \tablenotetext{b}{Synthesized FWHM beam size with robust
weighting}
\end{deluxetable}


\begin{deluxetable}{lcccccc}
\tabletypesize{\scriptsize}\tablecaption{\footnotesize Parameters
from N$_2$H$^+$ (1-0) spectral fitting$^{a}$\label{tbl-3}}
\tablewidth{0pt}

\tablehead{\colhead{Source} &\colhead{$V_{LSR}$}
&\colhead{$T_{ex}$} &\colhead{$\tau_{tot}$}
&\colhead{$\triangle$$v$} &\colhead{$\triangle$$v_{mean}$$^b$}
&\colhead{$\triangle$$v_{NT}$$^c$}\\
\colhead{Name} &\colhead{(km s$^{-1}$)} &\colhead{(K)} &\colhead{}
&\colhead{(km s$^{-1}$)} &\colhead{(km s$^{-1}$)} &\colhead{(km
s$^{-1}$)} } \startdata

IRAS\,03282  & 6.89$\pm$0.01     & 4.98$\pm$0.05   &  1.38$\pm$0.08  & 0.51$\pm$0.03 & 0.29$\pm$0.01 & 0.26$\pm$0.02\\
IRAS\,04166  & 6.64$\pm$0.02     & 4.08$\pm$0.13   &  1.76$\pm$0.18  & 0.34$\pm$0.01 & 0.32$\pm$0.01 & 0.29$\pm$0.02\\
RNO\,43      & 9.77$\pm$0.01     & 5.15$\pm$0.25   &  0.77$\pm$0.01  & 0.41$\pm$0.02 & 0.39$\pm$0.01 & 0.37$\pm$0.02\\
CB\,68       & 5.26$\pm$0.04     & 4.25$\pm$0.04   &  1.38$\pm$0.09  & 0.37$\pm$0.04 & 0.33$\pm$0.01 & 0.30$\pm$0.02\\
L723\,VLA2   & 11.00$\pm$0.02    & 5.74$\pm$0.21   &  0.78$\pm$0.03  & 1.07$\pm$0.04 & 0.51$\pm$0.01 & 0.49$\pm$0.02\\
CB\,188      & 7.23$\pm$0.02     & 4.58$\pm$0.08   &  0.73$\pm$0.04  & 0.38$\pm$0.03 & 0.31$\pm$0.01 & 0.28$\pm$0.02\\
CB\,224      & $-$2.65$\pm$0.01  & 4.61$\pm$0.04   &  1.75$\pm$0.08  & 0.50$\pm$0.03 & 0.47$\pm$0.01 & 0.45$\pm$0.02\\
CB\,230      & 2.78$\pm$0.01     & 5.49$\pm$0.13   &  1.54$\pm$0.04  & 0.36$\pm$0.01 & 0.29$\pm$0.01 & 0.26$\pm$0.02\\
CB\,244      & 3.24$\pm$0.01     & 4.93$\pm$0.11   &  1.09$\pm$0.16  & 0.38$\pm$0.02 & 0.45$\pm$0.02 & 0.43$\pm$0.03\\

\enddata
\tablenotetext{a}{Value at the intensity peak. The error
represents 1 $\sigma$ error in the hyperfine fitting.}
\tablenotetext{b}{Mean line width for each object (see
$\S$3.4).}\tablenotetext{c}{Non-thermal line width for each object
at the given gas temperature (10\,K; see $\S$3.4).}
\end{deluxetable}

\begin{deluxetable}{lcccccc}
\tabletypesize{\scriptsize} \tablecaption{\footnotesize Size,
density, and mass\label{tbl-4}} \tablewidth{0pt}
\tablehead{\colhead{Source} &\colhead{$R$} &\colhead{$M_{vir}$}
&\colhead{$\langle$$n_{H_2}$$\rangle$$^a$} &\colhead{$N(N_{2}H^{+})$} &\colhead{$M_{N_2H^+}$}&\colhead{$X(N_2H^+)$}\\
\colhead{}&\colhead{[AU]}&\colhead{[$M_{\odot}$]}&\colhead{[$\times$10$^5$
cm$^{-3}$]}&\colhead{[$\times$10$^{12}
$cm$^{-2}$]}&\colhead{[$\times$10$^{-10}$
$M_{\odot}$]}&\colhead{[$\times$10$^{-10}$]}}\startdata
IRAS\,03282  & 1703 & 0.42 & 22.6   & 1.81  & 1.57 & 5.11 \\
IRAS\,04166  & 1232 & 0.30 & 43.6   & 0.96  & 0.31 & 1.41 \\
RNO\,43 MM   & 3530 & 1.21 & 7.4    & 0.82  & 2.32 & 2.60 \\
CB\,68       & 1440 & 0.37 & 33.7   & 0.93  & 0.51 & 1.83 \\
L723\,VLA2   & 1272 & 0.65 & 84.4   & 2.84  & 1.89 & 3.96 \\
CB\,188      & 1763 & 0.44 & 21.6   & 0.57  & 0.46 & 1.42 \\
CB\,224      & 1329 & 0.57 & 64.7   & 1.95  & 2.07 & 4.97 \\
CB\,230      & 2762 & 0.71 & 9.0    & 1.52  & 3.86 & 7.41 \\
CB\,244      & 1774 & 0.72 & 34.9   & 0.96  & 0.69 & 1.30 \\
\enddata
\tablenotetext{a}{Mean density, computed from $R$ and $M_{vir}$
(see $\S$3.2)}
\end{deluxetable}

\begin{deluxetable}{lccccc}
\tabletypesize{\scriptsize} \tablecaption{\footnotesize Velocity
gradient and specific angular momentum \label{tbl-5}}
\tablewidth{0pt} \tablehead{\colhead{}&\colhead{mean
velocity}&\colhead{$g$}&\colhead{$\Theta_{g}$}&\colhead{$g_{r}$}
&\colhead{$J/M$}\\
\colhead{Source}&\colhead{[km s$^{-1}$]}&\colhead{[km s$^{-1}$
pc$^{-1}$]}&\colhead{[degree]}&\colhead{[km
s$^{-1}$]}&\colhead{[$\times$10$^{-3}$ km s$^{-1}$ pc]}}\startdata
IRAS\,03282  & 6.91 & $<$6.6        & 70.8$\pm$1.2   & 0.11  & $<$0.12  \\
IRAS\,04166  & 6.64 & $<$12.5       & -134.2$\pm$1.7 & 0.15  & $<$0.12  \\
RNO\,43 MM   & 9.72 & 5.8$\pm$0.1   & -23.1$\pm$1.0  & 0.21  & 0.45     \\
CB\,68       & 5.15 & $<$10.3$\pm$0.2  & 161.2$\pm$0.8  & 0.15  & $<$0.13  \\
L723\,VLA2   & 11.03& $<$24.2       & -139.2$\pm$0.2 & 0.30  & $<$0.24  \\
CB\,224      & -2.67&  $<$11.2      & -82.7$\pm$1.0  & 0.15  & $<$0.12  \\
CB\,230      & 2.69 & 8.8$\pm$0.1   & -54.0$\pm$0.4  & 0.24  & 0.42     \\
CB\,244      & 3.51 & $<$22.9  & 51.2$\pm$0.2   & 0.41  & $<$0.45  \\
\enddata
\end{deluxetable}

\begin{deluxetable}{lccccc}
\tabletypesize{\scriptsize} \tablecaption{\footnotesize Energy
balance \label{tbl-6}} \tablewidth{0pt}
\tablehead{\colhead{Source} &\colhead{$E_{grav}^a$ ($\times
10^{35}$$J$)} &\colhead{$\beta_{rot}^b$}
&\colhead{$\beta_{therm}^c$} &\colhead{$\beta_{turb}^d$}
&\colhead{$\beta_{vir}^e$}}\startdata

IRAS\,03282  & 1.34 &  $<$0.004  & 0.33 & 0.11  & $-$0.11  \\
IRAS\,04166  & 0.99 &  $<$0.007  & 0.33 & 0.14  & $-$0.05  \\
RNO\,43 MM   & 5.51 &  0.009     & 0.24 & 0.16  & $-$0.19  \\
CB\,68       & 1.29 &  $<$0.006  & 0.31 & 0.14  & $-$0.08  \\
L723\,VLA2   & 4.36 &  $<$0.014  & 0.16 & 0.19  & $-$0.27  \\
CB\,224      & 3.19 &  $<$0.004  & 0.19 & 0.19  & $-$0.23  \\
CB\,230      & 2.40 &  0.017     & 0.31 & 0.11  & $-$0.12  \\
CB\,244      & 3.91 &  $<$0.03   & 0.20 & 0.18  & $-$0.18  \\
\enddata
\tablenotetext{a}{Gravitational potential energy: $E_{grav}$ =
$\alpha_{vir}GM^2/R$.} \tablenotetext{b}{$\beta_{rot}$ =
$\frac{E_{rot}}{E_{grav}}$ =
$\frac{1}{2}\frac{\alpha_{rot}}{\alpha_{vir}}\frac{\omega^2R^3}{GM}$
= 0.17$\frac{g^2}{sin^2i} \frac{R^3}{GM}$. Here we assume $sini =1
$ for all the objects.}\tablenotetext{c}{$\beta_{therm}$ =
$\frac{E_{therm}}{E_{grav}}$ and $E_{therm}$ is the thermal energy
estimated from $E_{therm}$ = $\frac{3}{2} M \frac{kT}{\mu m_{H}}$
($k$ is the Boltzman constant and $\mu$ = 2.33 is the mean
molecular weight).}\tablenotetext{d}{$\beta_{turb} =
\frac{E_{turb}}{E_{grav}}$ and $E_{turb}$ is the turbulent energy
estimated from the non-thermal line width $E_{turb} =
3/2M\sigma_{NT}^2$ ($\sigma_{NT}^2$ = $\frac{\triangle
v_{turb}}{8ln2}$. Here we assume that the non-thermal line widths
at the given temperature in our sample are from
turbulence.}\tablenotetext{e}{$\beta_{vir}$ = 2($\beta_{rot}$ +
$\beta_{therm}$ + $\beta_{turb}$) $-$ 1.}

\end{deluxetable}


\clearpage
\begin{figure*}
\begin{center}
\includegraphics[width=16cm,angle=0]{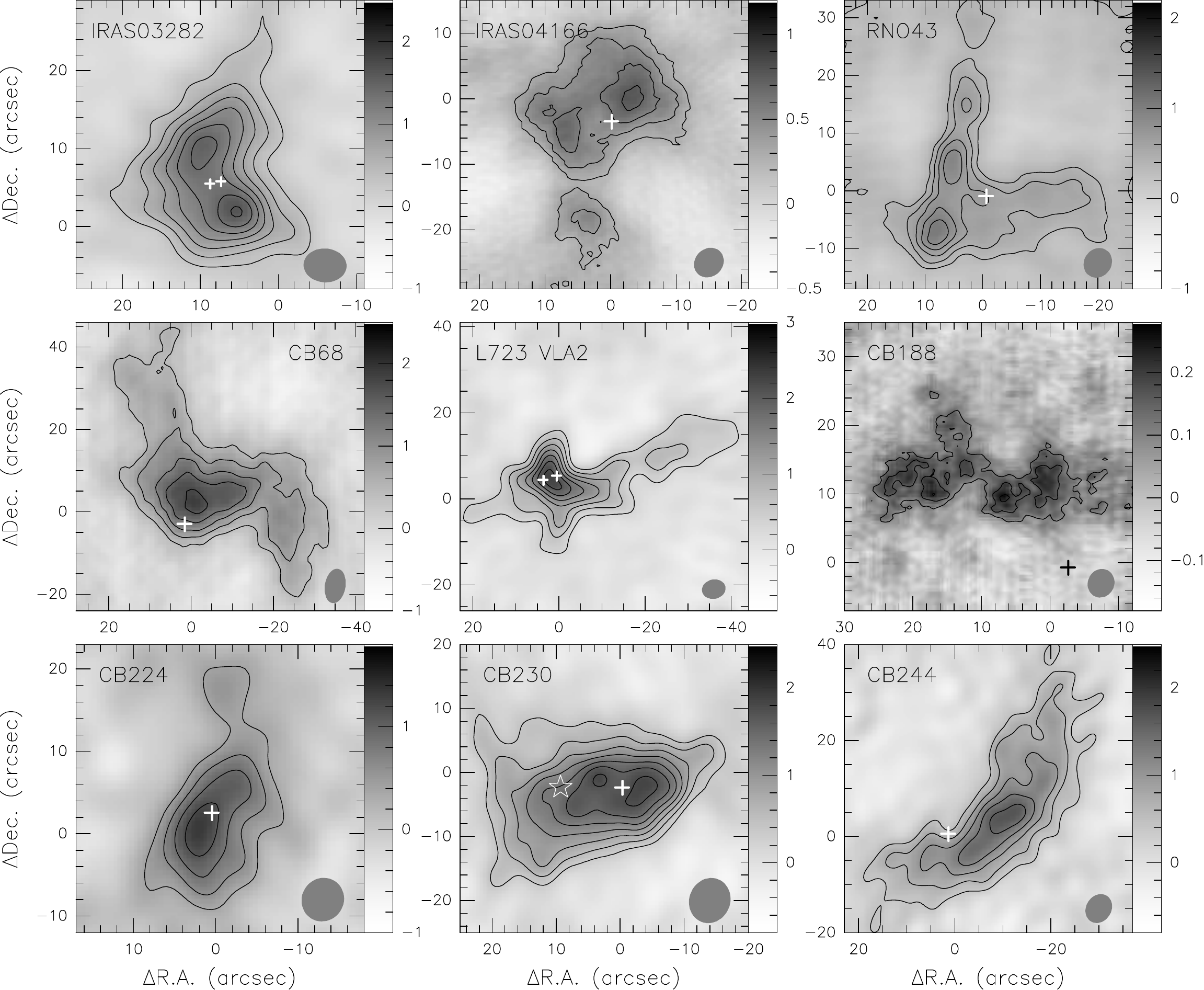}
\caption{Maps of the N$_2$H$^+$ (1$-$0) intensity integrated over
the seven hyperfine components for 9 protostars. The unit of the
scale is [Jy beam$^{-1}$ km s$^{-1}$]. The contours start at
$\sim$ 3~$\sigma$ with steps of $\sim$ 2~$\sigma$. Beam sizes are
shown as grey ovals in each map. The crosses in the maps represent
the peaks of 3\,mm dust continuum emission (except IRAS\,03282, in
which the crosses indicate the peaks of 1.3\,mm dust emission).
All the positions of dust emission are selected from Launhardt et
al. 2007 (in prep.). The asterisk in CB\,230 marks the positions
of the secondary protostar observed at 7\,$\mu$m with ISOCAM (not
detected at 3\,mm, see Launhardt 2001). \label{fig1}}
\end{center}
\end{figure*}

\clearpage
\begin{figure*}
\begin{center}
\includegraphics[width=16cm,angle=0]{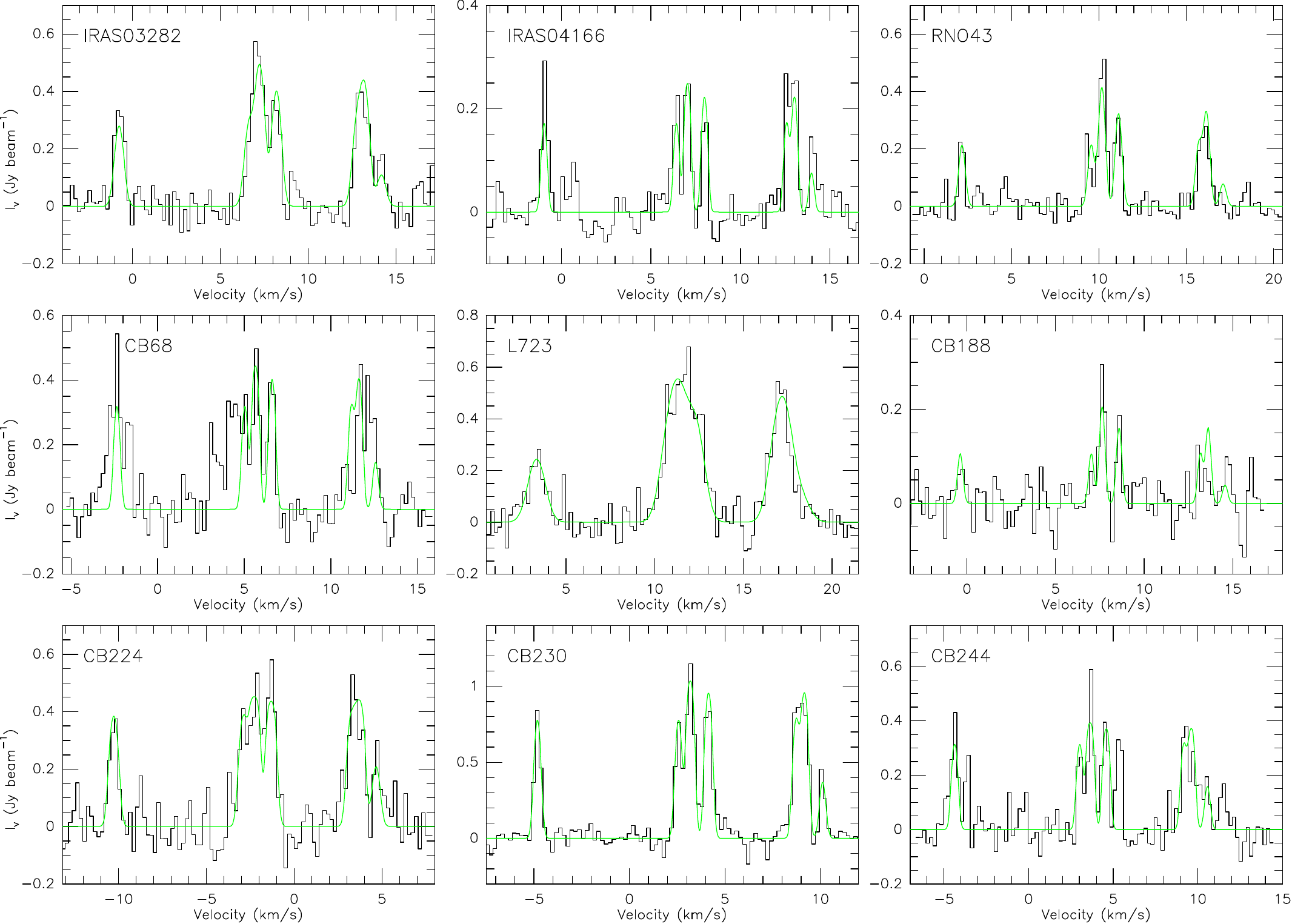}
\caption{N$_2$H$^+$ spectra at the peak position of the nine
observed protostars. Gray dotted curves show the hyperfine
structure line fitting. Fitting results are given in
Table~3.\label{fig2}}
\end{center}
\end{figure*}

\clearpage
\begin{figure*}
\begin{center}
\includegraphics[width=16cm,angle=0]{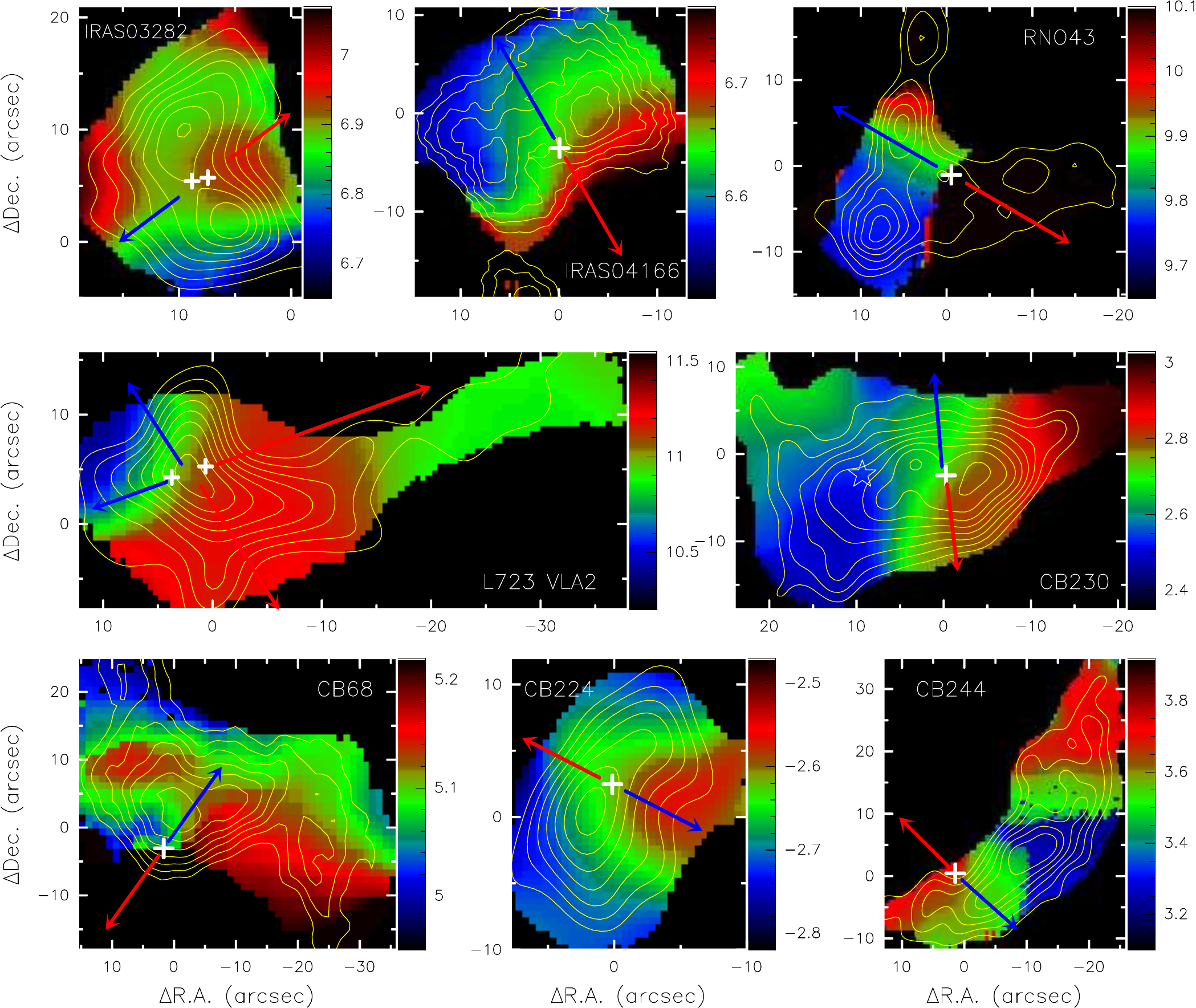}
\caption{N$_2$H$^+$ velocity field maps of 8 Class 0 protostars.
The unit of the scale is km s$^{-1}$. The contours in each map are
from Fig.\,1, but range from 30\% to 99\% of the peak intensity by
the step of 10\%. The crosses in each are same as them in Fig.\,1.
The red and blue arrows show the directions of the red- and
blue-shifted outflow for each source (see text).\label{fig3}}
\end{center}
\end{figure*}

\clearpage

\begin{figure*}
\begin{center}
\includegraphics[width=16.0cm, angle=0]{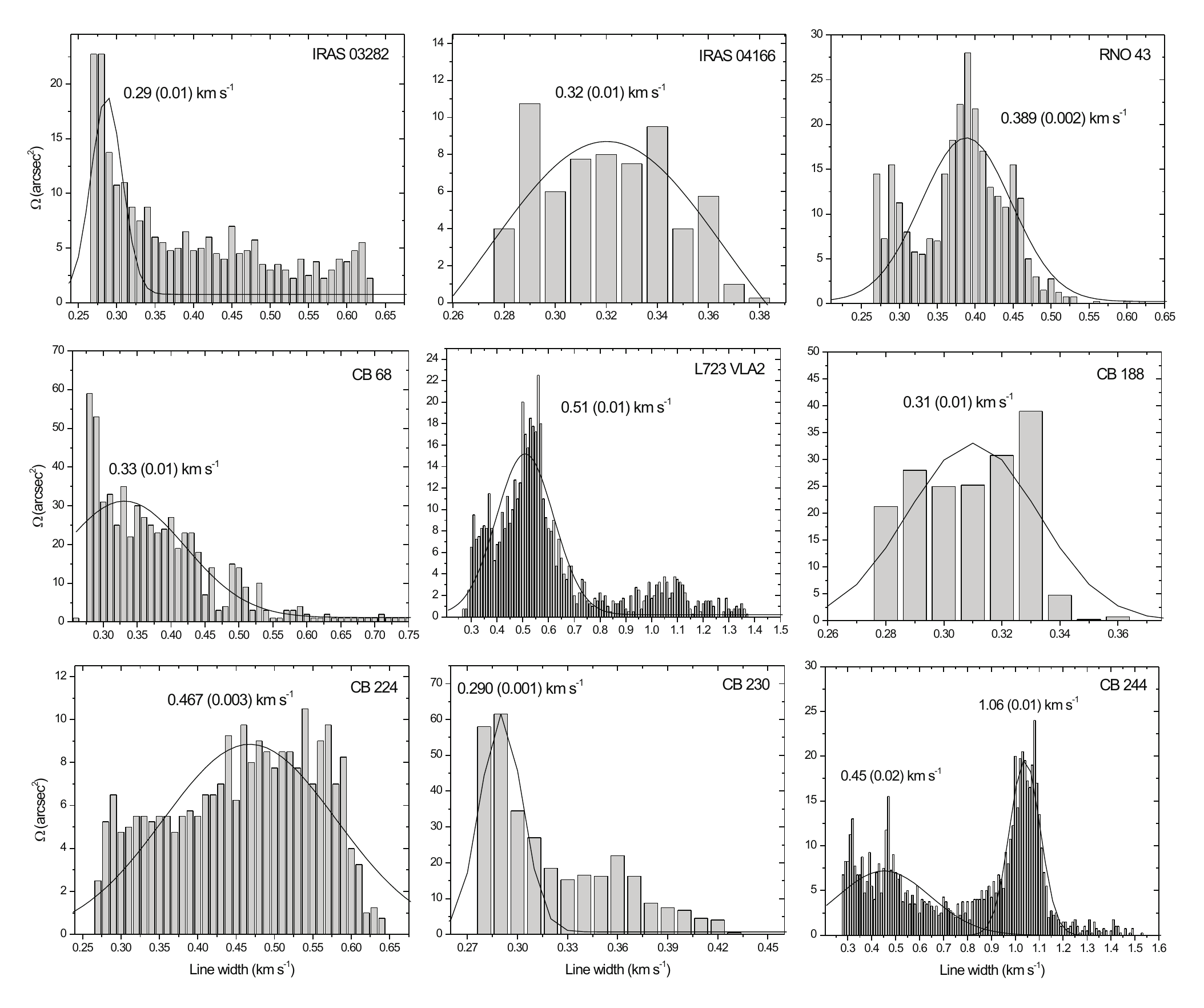}
\caption[]{Diagrams of the correlation between line width and
solid angle area for the nine protostars in our sample.
Black solid curves and values in each map show the results of
Gaussian fitting.\label{fig4}}
\end{center}
\end{figure*}

\clearpage
\begin{figure*}
\begin{center}
\includegraphics[width=16cm, angle=0]{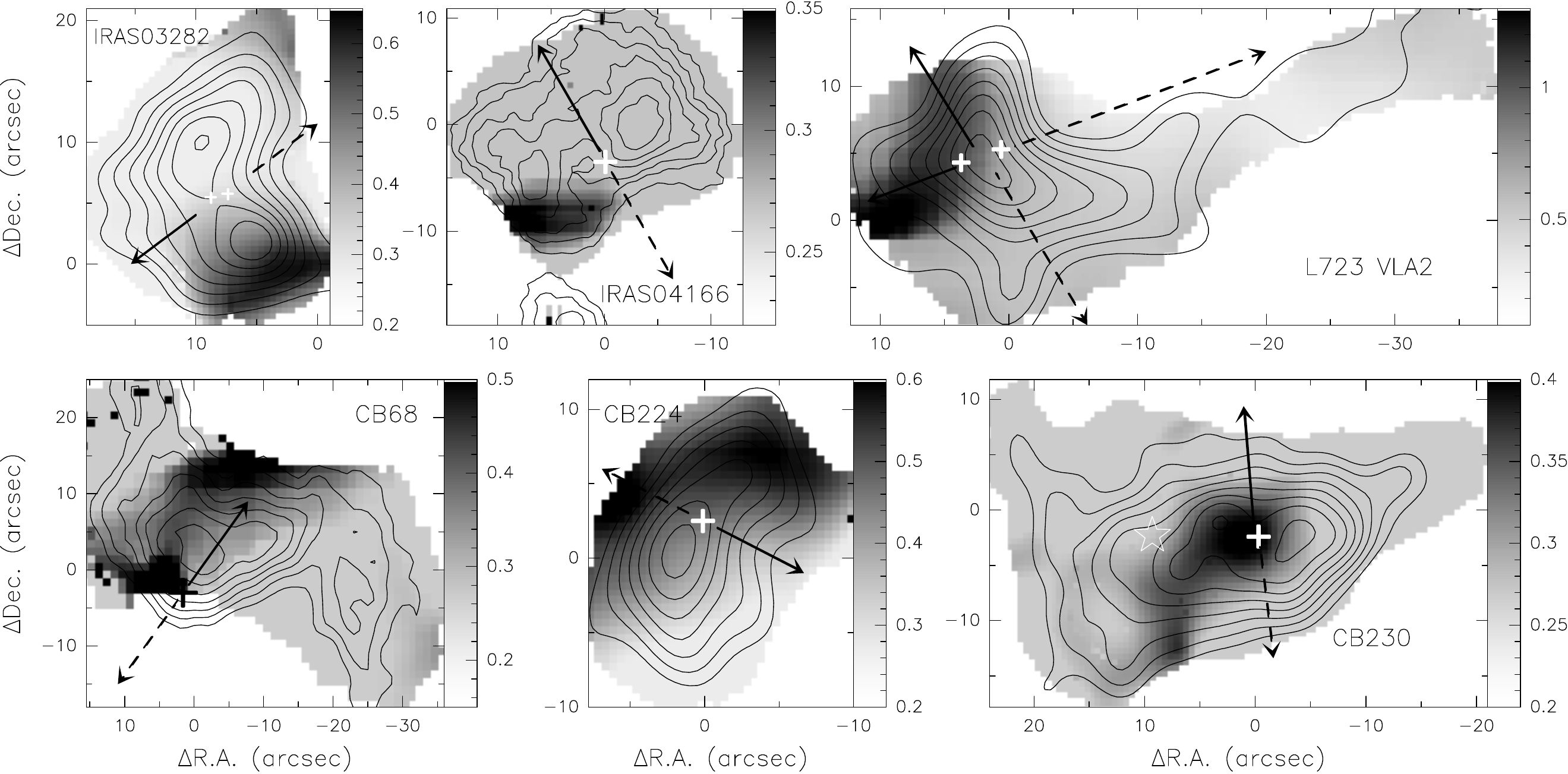}
\caption{Line widths distribution of 6 Class 0 protostars in our
sample. The unit of the scale is km s$^{-1}$. The crosses in each
map are same as them in Fig.\,1 and the contours in each map are
same as them in Fig.\,3. Solid and dashed lines in each map show
the direction of blue- and red-shifted outflow,
respectively.\label{fig5}}
\end{center}
\end{figure*}

\clearpage
\begin{figure*}
\begin{center}
\includegraphics[width=14cm, angle=0]{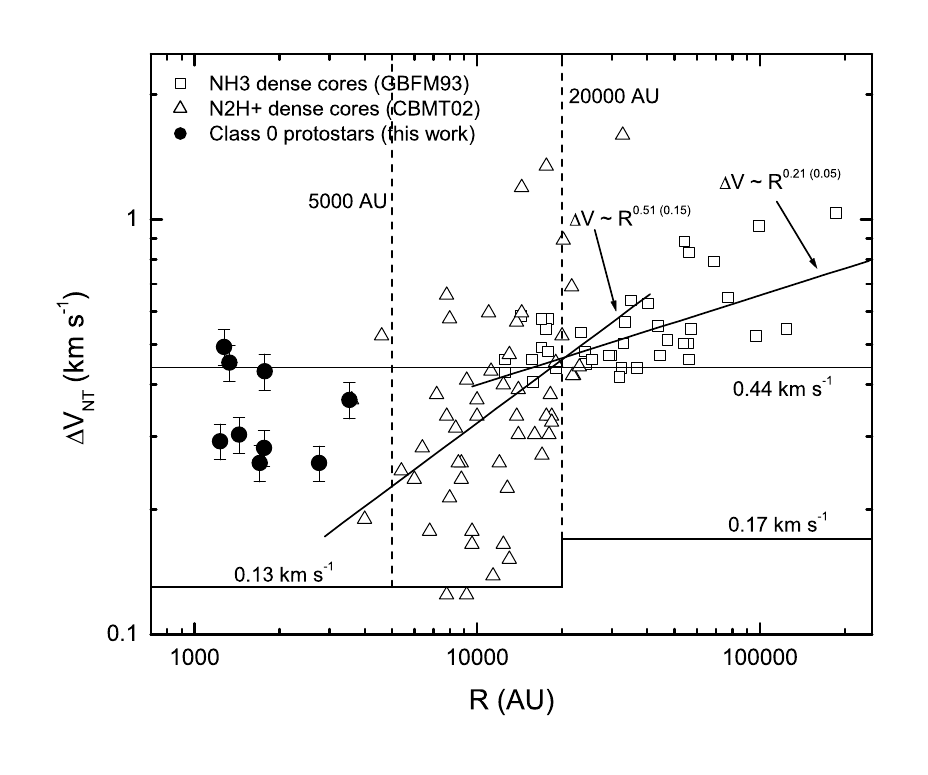}
\caption{Non-thermal line width $\triangle$V$\rm_{NT}$ vs. size
$R$ of dense molecular cloud cores. Data of NH$_{3}$ (open
squares) and N$_{2}$H$^{+}$ (open triangles) dense cores are
adopted from GBFM93 and CBMT02, respectively. Solid lines marked
0.13, 0.17 and 0.44 km s$^{-1}$ represent thermal line widths of
N$_{2}$H$^{+}$, NH$_{3}$, and 2.33\,m$\rm_H$ mass ``average"
particle at 10\,K, respectively. The fit to the GBFM93 and
CBMT02 data results in power-law indexes of 0.21$\pm$0.05 and 
0.51$\pm$0.15, respectively. The levels (p-values) of statistical 
significance are $<$ 0.01\% (GBFM93) and $\sim$ 0.1\% (CBMT02),
respectively. \label{fig6}}
\end{center}
\end{figure*}

\clearpage
\begin{figure*}
\begin{center}
\includegraphics[width=14cm, angle=0]{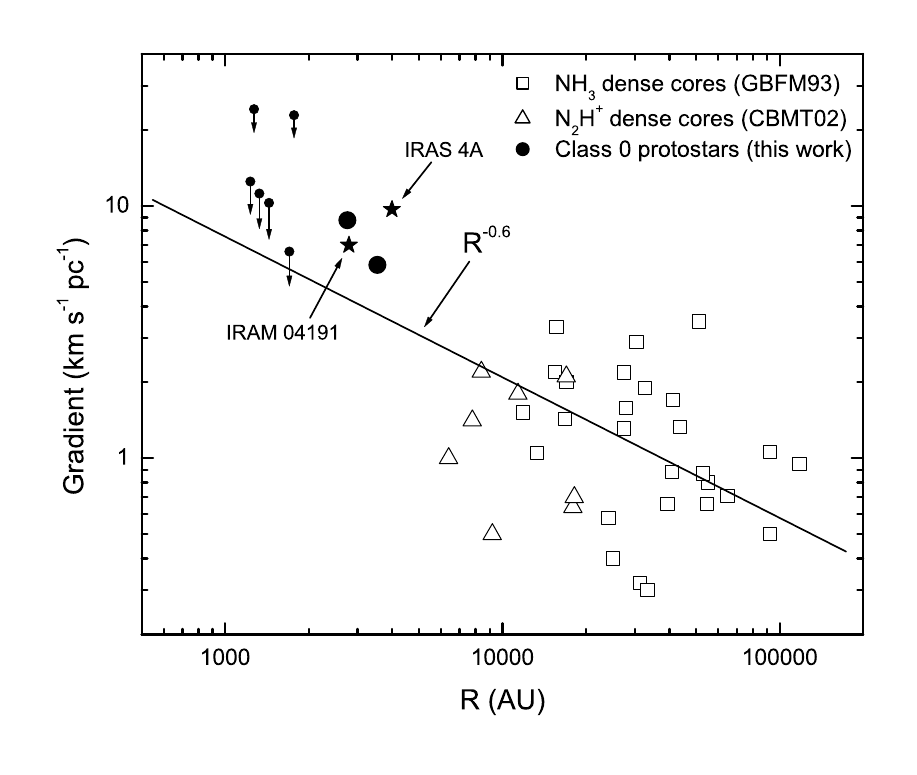}
\caption{Mean velocity gradient vs. size $R$ of dense molecular
cloud cores. Data of NH$_{3}$ (open squares) and N$_{2}$H$^{+}$
(open triangles) dense cores are adopted from GBFM93 and CBMT02,
respectively. Data of IRAM\,04191 and NGC1333 IRAS4A (asterisks)
are adopted from Belloche et al. (2002; 2006). Solid line shows
the fitting result with a power-law index of $-$0.6$\pm$0.1. The
level of statistical significance is $\sim$ 0.1\%. \label{fig7}}
\end{center}
\end{figure*}

\clearpage
\begin{figure*}
\begin{center}
\includegraphics[width=14cm, angle=0]{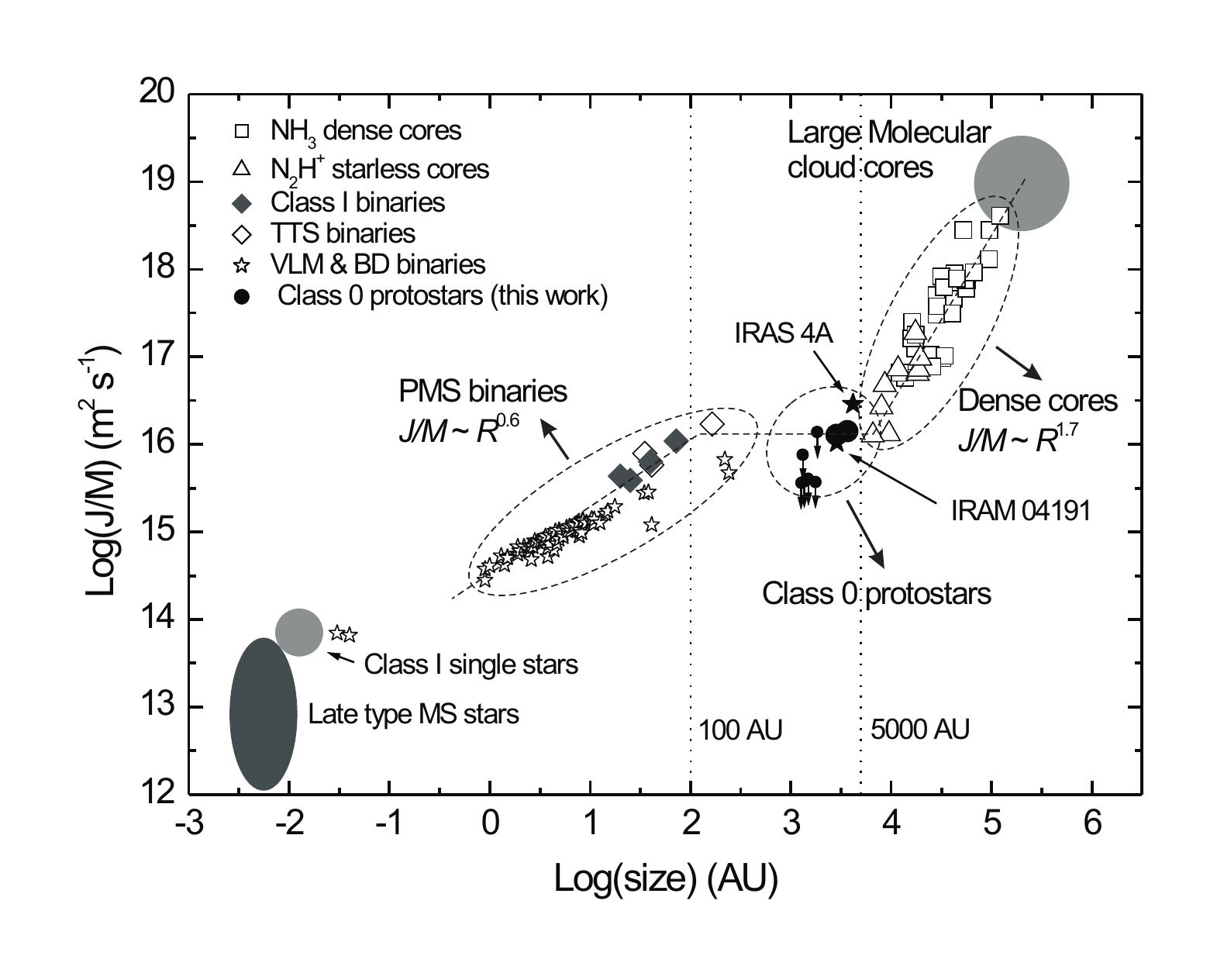}
\caption{Distribution of specific angular momentum $J/M$ vs. size
$R$ of molecular cloud cores, protostars, and stars. For the dense
cores, the specific angular momenta are derived from $J/M$ = 2/5
$(R^{2} \times g$) (see text). The data of the NH$_{3}$ dense
cores are from Goodman et al. (1993); the data of the
N$_{2}$H$^{+}$ starless cores are from Caselli et al. (2002). For
the binary stars, the orbital specific angular momenta are derived from
$J/M$ = $\sqrt{GMD}$$\times q/(1+q)^2$ (see text). The data of
Class I binaries and T Tauri binaries are from Chen et al.
(2007\,IV, in prep.), and the masses are all dynamic masses; The
data of very low-mass ($<$ 0.1 $M_\odot$) and brown dwarf binaries
are from Burgasser et al. (2007).\label{fig8}}
\end{center}
\end{figure*}

\clearpage
\begin{figure*}
\begin{center}
\includegraphics[width=14cm, angle=0]{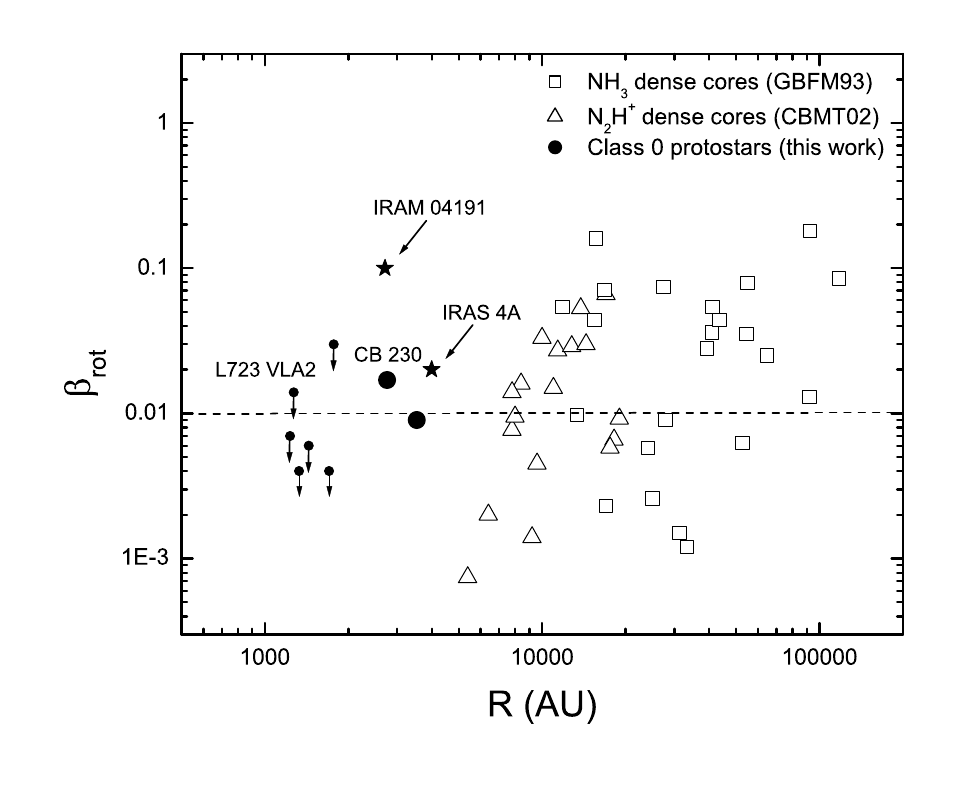}
\caption{Ratio of rotational to gravitational energy $\beta_{rot}$
vs. size $R$. Data of NH$_{3}$ (open squares) and N$_{2}$H$^{+}$
(open triangles) dense cores are adopted from GBFM93 and CBMT02,
respectively. Data of IRAM\,04191 and NGC1333 IRAS4A (asterisks)
are adopted from Belloche et al. (2002; 2006).\label{fig9}}
\end{center}
\end{figure*}

\end{document}